\documentclass[conference]{IEEEtran}
\IEEEoverridecommandlockouts
\usepackage{cite}
\usepackage{amsmath,amssymb,amsfonts}
\usepackage{algorithmic}
\usepackage{graphicx}
\usepackage{textcomp}
\usepackage{xcolor}
\usepackage{multirow}
\usepackage{threeparttable}
\usepackage{url}

\def\BibTeX{{\rm B\kern-.05em{\sc i\kern-.025em b}\kern-.08em
    T\kern-.1667em\lower.7ex\hbox{E}\kern-.125emX}}
\begin{document}

\title{A High-Throughput Hardware Accelerator for Lempel-Ziv 4 Compression Algorithm\\}

\author{Tao Chen\textsuperscript{1}, Suwen Song\textsuperscript{2}, and Zhongfeng Wang\textsuperscript{1,2} \textit{Fellow, IEEE}\\
	
	\textsuperscript{1} School of Electronic Science and Engineering, Nanjing University, Nanjing, China\\
	\textsuperscript{2} School of Integrated Circuits, Sun Yat-sen University, Shenzhen, China\\
	Email: taochen@smail.nju.edu.cn, songsw5@mail.sysu.edu.cn, zfwang@nju.edu.cn 
	\vspace{-1.2em}
}

\maketitle

\begin{abstract}

This paper delves into recent hardware implementations of the Lempel-Ziv 4 (LZ4) algorithm, highlighting two key factors that limit the throughput of single-kernel compressors. Firstly, the actual parallelism exhibited in single-kernel designs falls short of the theoretical potential. Secondly, the clock frequency is constrained due to the presence of the feedback loops. To tackle these challenges, we propose a novel scheme that restricts each parallelization window to a single match, thus elevating the level of actual parallelism. Furthermore, by restricting the maximum match length, we eliminate the feedback loops within the architecture, enabling a significant boost in throughput. Finally, we present a high-speed hardware architecture. The implementation results demonstrate that the proposed architecture achieves a throughput of up to 16.10 Gb/s, exhibiting a 2.648$\times$ improvement over the start-of-the-art. The new design only results in an acceptable compression ratio reduction ranging from 4.93\% to 11.68\% with various numbers of hash table entries, compared to the LZ4 compression ratio achieved by official software implementations disclosed on GitHub.

\end{abstract}

\begin{IEEEkeywords}
LZ4, lossless compression, FPGA, high throughput, hardware accelerator.
\end{IEEEkeywords}

\section{Introduction}

In the era of big data, massive amounts of data pose significant challenges to transmission bandwidth, data storage, and data processing. As an effective method to reduce data size, lossless compression algorithms have been widely used in most data application scenarios\cite{kavitha2016survey,ferraro2021fasta,jia2020slimcache}. For instance, the Lempel-Ziv 77 (LZ77)\cite{ziv1977universal} algorithm and its variant, Lempel-Ziv-Stac (LZS) algorithm are utilized for compressing IP packets to reduce network transmission overhead.

Most compression algorithms were originally implemented in software, resulting in relatively slow speeds. However, with the rapid development of  digital technologies, such as big data and multimedia, the demand of high-throughput implementation for compression algorithms is steadily growing. Consequently, there is a rising interest in transferring these compression tasks from software platforms to hardware solutions, in which field-programmable gate arrays (FPGAs) have emerged as a popular choice.

The LZ4 algorithm, introduced by Collet\cite{collet2011real} in 2011, is a variant of LZ77. Known for its outstanding compression speed, LZ4 surpasses other LZ algorithms in implementations. Besides, a comprehensive  analysis of LZ4 from the perspective of hardware design\cite{bartik2015lz4} has been conducted, revealing additional benefits of the LZ4 algorithm. These include low latency and high resource efficiency, making it a highly advantageous choice for hardware-based compression tasks. Hence, we select LZ4 as a candidate for our design in this paper.

Several FPGA-based acceleration designs for LZ4 compression have been proposed\cite{bartik2015lz4,liu2018data,lee2017design,bartik2019design,benevs2019high,liu2022hybridc}. In 2015, Bartik \textit{et al.} \cite{bartik2015lz4} pioneered the use of an FPGA to implement LZ4 for compressing 4K transmissions. Liu \textit{et al.}\cite{liu2018data} refined their work and introduced a modified LZ4 format to enhance the clock frequency. In order to further improve throughput, in 2019 Bartik and Benes \textit{et al.}\cite{bartik2019design,benevs2019high} utilized parallelism in the match searching stage, resulting in a significant increase in throughput. Based on the parallel implementation, \cite{liu2022hybridc} further proposed a scheme aiming at minimizing resource utilization to enable the deployment of multiple kernels within the same resource constraints, leading to an enhancement in throughput.

While some existing architectures\cite{lee2017design,bartik2019design,benevs2019high,liu2022hybridc} utilized parallelism to improve throughput, they avoid sacrificing compression ratio by traversing all non-overlapping matches in the extended match stage. This results in a significant reduction in parallelism during the extended match stage, which fails to match the parallelism achieved in previous stages, consequently greatly reducing the throughput.
Besides, in the current parallel hardware architectures\cite{bartik2019design,benevs2019high,liu2022hybridc}, the presence of timing dependencies in the address signals during the extended match stage forms the feedback loops, which inhibits further increment of frequency through pipeline insertion, thereby limiting the upper bound of throughput.

The objective of this paper is to address the aforementioned issues to improve throughput.
To tackle the mismatch in parallelism, we propose a solution where each parallelization window is constrained to contain a single match, ensuring that each component of the architecture exhibits the same level of parallelism.
In order to address the loops in the architecture, we aim to establish a maximum limit on the match length during the extended match stage. This approach ensures that the architecture becomes fully feedforward, enabling the insertion of pipelines to enhance frequency. Furthermore, it guarantees predictable output delays, albeit at the cost of a certain degree of compression ratio loss. Given certain scenarios, such as multimedia applications, where the significance of the throughput and predictability outweighs that of compression ratio\cite{bartik2015lz4,bartik2020sight}, sacrificing compression ratio becomes acceptable.
The implementation of our proposed architecture on FPGA demonstrates that our throughput is up to 16.10 Gb/s, which is 2.648$\times$ higher than that of the best-performing architecture in the open literatures.

The rest of the paper is structured as follows. Section II provides a review of relevant works and investigates the throughput bottlenecks in current parallel architectures. In Section III, we introduce solutions to mitigate throughput bottlenecks while analyzing the impact on compression ratio. The hardware architecture of our proposed scheme and a comparison with other FPGA designs of LZ4 algorithm are presented in Section IV. Finally, Section V concludes the paper.

\section{Preliminaires}

\subsection{LZ4 Algorithm}
\begin{figure}[t]
	\centerline{\includegraphics[width=\linewidth]{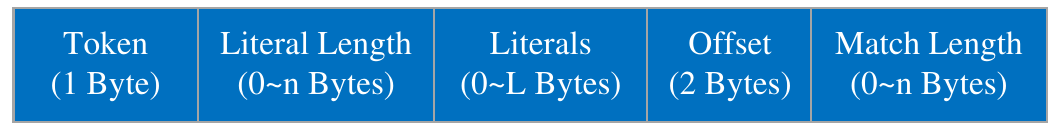}}
	\caption{Data format of an LZ4 sequence.}
	\label{fig1}
\end{figure}

LZ4 is a dictionary-based compression algorithm\cite{sayood2017introduction}, which was initially conceived as a distinct compressed data format. Compressed data files are composed of LZ4 sequences, which include a token, literal length, literals, offset, and match length,  as depicted in Fig.~\ref{fig1}.

LZ4 achieves data compression by leveraging statistical redundancy in data patterns and employing a mechanism to eliminate repeated literals. It identifies recurring data strings from previous data and substitutes them with an offset-length pair, utilizing a dictionary to determine the offset of previous literals and their match length.

\subsection{Parallel Implementation of the LZ4 Algorithm}
\begin{figure}[t]
	\centerline{\includegraphics[width=\linewidth]{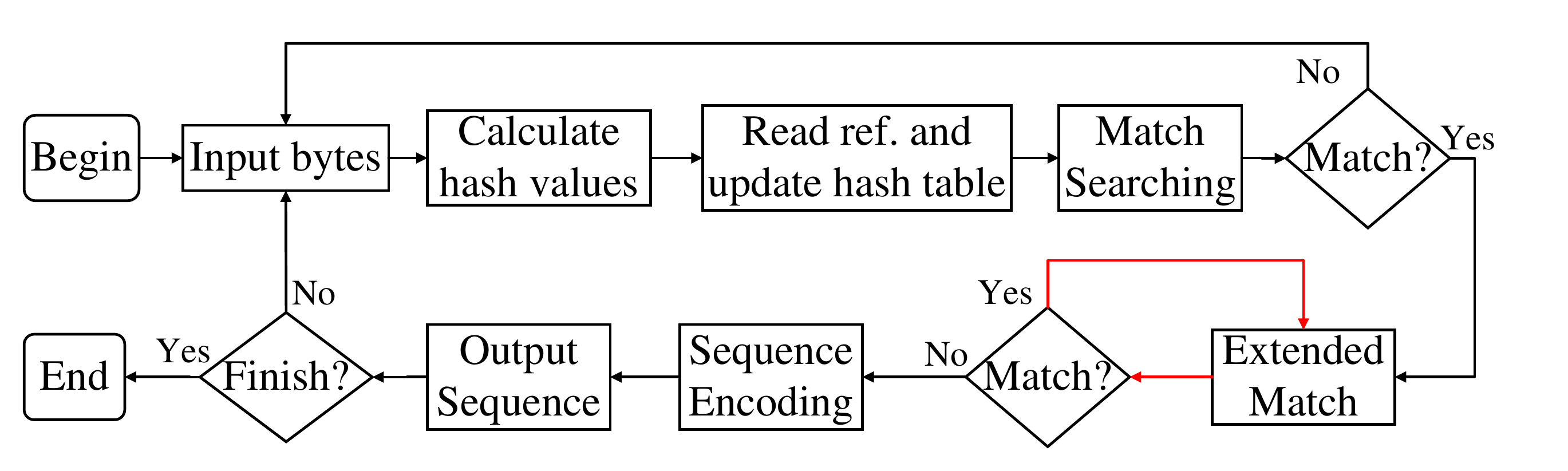}}
	\caption{The flow chart of LZ4 algorithm.}
	\label{fig2}
	\vspace{-2mm}
\end{figure}

A typical hardware parallel implementation of the LZ4 compression algorithm consists of five major stages, including hash calculation, hash table update, match searching, extended match, and sequence encoding, as illustrated in Fig.~\ref{fig2}. We use a designed \textit{Parallelization Window Size (PWS)} to  represent parallelism. Within this parallelization window, a set of spatial locations enables concurrent execution of the compression algorithm across multiple consecutive bytes.
 
\paragraph{Hash Calculation} The Fibonacci hashing principle, known for its simplicity and efficiency achieved through straightforward multiplication with a constant 2654435761, is widely adopted as the most common methods for hash calculation algorithms\cite{bartik2015lz4,liu2018data,lee2017design,bartik2019design,benevs2019high,liu2022hybridc}. This 32-bit constant can be efficiently multiplied by four DSP48 slices available on the FPGA. The generated IP (Intellectual Property) core is pipelined, and the multiplication result is truncated to extract the most significant bits used for the hash table address.

\paragraph{Hash Table Update} LZ4 utilizes a hash table to swiftly identify repeated data with a constant time complexity of $\mathcal{O}$(1). Indexed by the hash value of current string, the hash table retrieves the candidate address and four bytes of the candidate string, where the length of four bytes is the minimum match length specified by the LZ4 algorithm. Subsequently, the address of the current string and its corresponding four-byte data are stored, replacing the previous ones\cite{liu2018data,bartik2019design,benevs2019high}.

\paragraph{Match Searching} The LZ4 algorithm compares the four bytes of the candidate string with the current four bytes to validate the match. To identify the longest match, the earliest candidate among \textit{PWS} possible matches should be selected, which corresponds to the input string located near the beginning of the parallelization window. Upon successful matching between the candidate and current strings, an extended match process is initiated to search for a longer match. Conversely, if no match is found, the input window advances by \textit{PWS} bytes, and the process repeats.

\paragraph{Extended Match} Upon detecting a 4-byte match, we proceed by fetching \textit{PWS} bytes data from the address obtained by advancing 4 bytes from the candidate address, and compare it with the current data. Depending on the result of this comparison, we determine whether to continue reading \textit{PWS} bytes from the address obtained by advancing (\textit{PWS} + 4) bytes from the candidate address in the following clock cycle, or to read \textit{PWS} bytes from a new address. This process repeats iteratively until the maximum match length is found.

For example, assume that the \textit{PWS} is set to 8 bytes, the current string is “abcdefghijklmnopqrst” and the candidate string is “abcdefghijklmnoheshe”. Upon discovering a 4-byte substring “abcd” match, we proceed to compare the next 8 bytes “efghijkl”. If the match length is equal to the window size, then we backtrack to find an 8-byte substring “mnoheshe”, and we find that only 3 bytes match. The match length is less than the \textit{PWS} bytes, which means the maximum match length is found.

Due to the temporal logic dependency, this process inevitably creates a feedback loop in the circuit\cite{bartik2019design,benevs2019high,liu2022hybridc}, which becomes a bottleneck for increasing the frequency. As illustrated in Fig.~\ref{fig2}, this loop is highlighted by the red line. 

Besides, in some implementations, the compression ratio is optimized by traversing all non-overlapping matches in the extended match stage, which can be achieved either by utilizing FIFOs to segregate the match searching module and the extended match module\cite{benevs2019high} or by advancing the starting address of the subsequent parallelization window to the end of the ongoing match\cite{liu2022hybridc}. However, this adjustment inherently diminishes the effective parallelism level of the compression kernel in comparison to the theoretical parallelism represented by \textit{PWS}. This occurs because, during the extended match stage, the sliding step becomes narrower than the \textit{PWS}, ultimately restricting the overall throughput.

\paragraph{Sequence Encoding} Upon completion of a match round, a comprehensive compression data set is produced, comprising literals, literal length, and details regarding matches (length and offset). Subsequently, the output encoding module encodes this information into sequences as shown in Fig.~\ref{fig1}.

\subsection{Summary}
We have conducted a thorough analysis of the typical LZ4 implementation on FPGA and identified two primary factors that significantly limit the potential for throughput enhancement in parallel architectures.
\begin{itemize}
\item Some implementations employ FIFO structures\cite{benevs2019high} or advance the parallelization window\cite{liu2022hybridc} to capture all non-overlapping matches, aiming to enhance the compression ratio. However, this approach inherently leads to a decrease in the effective degree of parallelism compared to the theoretical parallelism.

\item The presence of a loop within the current parallel architectures\cite{bartik2019design,benevs2019high,liu2022hybridc} poses as a bottleneck, constraining the operating frequency and subsequently limiting the maximum achievable throughput.

\end{itemize}

\section{The Proposed High-Throughput Compressor}
We propose solutions aimed at addressing the two aforementioned disadvantageous factors that constrain throughput and evaluate their influence on the compression ratio.

\subsection{Enhancement of Parallelism}

As previously discussed in Section II-B, researchers have employed FIFO or advanced the starting address of the next parallelization window to enhance the compression ratio, at the expense of reducing the parallelism. The rationale behind these operations ensuring a higher compression ratio is their ability to prevent the oversight of scenarios where multiple non-overlapping matches occur within the parallelization window.

\begin{figure}[t]
	\centerline{\includegraphics[width=\linewidth]{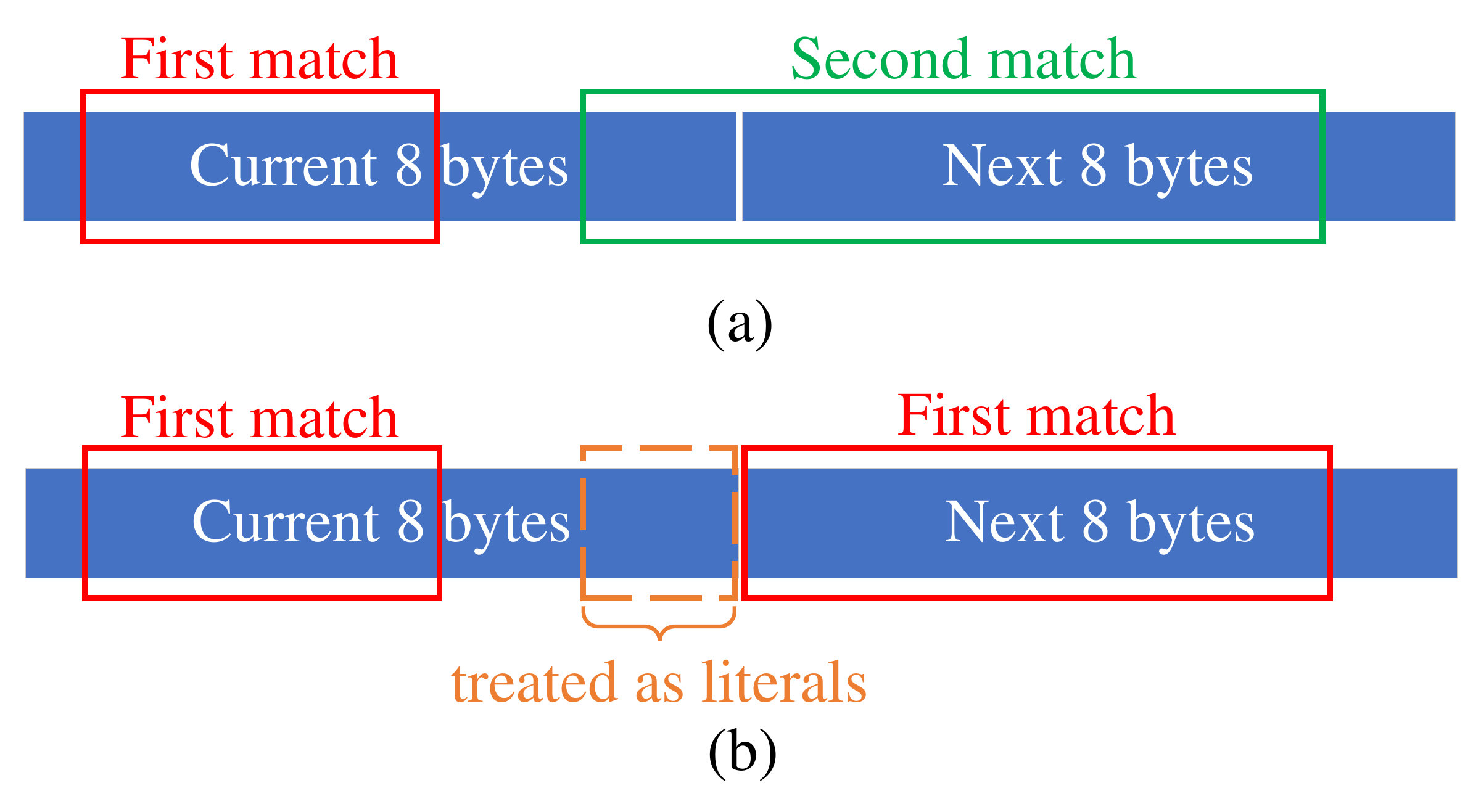}}
	\caption{The example of different schemes within current parallelization window.}
	\label{fig3}
\end{figure}

For instance, considering the case where \textit{PWS} is set to 8 bytes, there is only one scenario where multiple non-overlapping matches exist within the parallelization window. As depicted in Fig.~\ref{fig3}(a), the first match ends within the current parallelization window and the starting position of the subsequent match exists within the remaining data of the same window. 
In this situation, after spending a clock cycle to compress the first match, the compression of the second match within the same parallelization window still requires an additional clock cycle to be carried out, which results in a lower degree of parallelism compared to the previous compression stages where one parallelization window could be fully processed within a single clock cycle. 
So when encountering the scenario depicted in Fig.~\ref{fig3}(a), the actual parallelism will fall below the theoretical parallelism. The extent of this reduction is dependent on the frequency of occurrences of scenarios similar to Fig.~\ref{fig3}(a) within the dataset. 
This inevitable decrease in actual parallelism will ultimately result in a decrement in overall throughput. For example, Benes \textit{et al.}\cite{benevs2019high} suffered from a throughput reduction from 10 Gbps to 6.08 Gbps with the implementation of FIFO, while Liu \textit{et al.}\cite{liu2022hybridc} saw their throughput decrease from 6.4 Gbps to 4.5 Gbps by advancing the starting address of the subsequent parallelization window. Their throughput all decreased by approximately 30\% to 40\% due to the decline in parallelism.

Hence, to enhance the parallelism and align it with the theoretical parallelism, we constrain each parallelization window to process only a single match, as depicted in Fig.~\ref{fig3}(b). This approach does not continue to search for matches within the remaining data of the current window, and instead, it starts to find matches from the beginning of the next parallelization window, which will inevitably result in some data within the matched segment being treated as literals, as marked within the orange box in Fig.~\ref{fig3}(b), leading to a slight but acceptable compromise in compression ratio.

Subsequently, we evaluate the degradation in compression ratio resulting from our proposed approach using standard test datasets, commonly referred to as corpus. The corpus sets typically encompass a diverse range of input data, ensuring relevance to real-world scenarios. The Calgary corpus\cite{bell1989modeling}, is selected in the compression ratio evaluation. The performance of the aforementioned method is presented in Table~\ref{tab1}.

\begin{table}[t]
	\caption{Compression Ratio$^{\mathrm{a}}$ of Different Schemes within Parallelization Window on the Calgary Corpus}
	\vspace{-4mm}
	\begin{center}
		\resizebox{\linewidth}{!}{
			\begin{threeparttable}
			\renewcommand{\arraystretch}{1.5} 
			\begin{tabular}{|c|c|c|c|c|c|c|c|c|}
				\hline
				\multirow{2}{*}{\textbf{Scheme$^{\mathrm{b}}$}}& \multicolumn{8}{|c|}{\textbf{Number of entries in the hash table}} \\
				\cline{2-9} 
				& \textbf{64}& \textbf{128}& \textbf{256}& \textbf{512}& \textbf{1024}& \textbf{2048}& \textbf{4096}& \textbf{8192} \\
				\hline
				\multirow{2}{*}{\textbf{Multiple matches}}& \multirow{2}{*}{1.277}& \multirow{2}{*}{1.340}& \multirow{2}{*}{1.419}& \multirow{2}{*}{1.512}& \multirow{2}{*}{1.615}& \multirow{2}{*}{1.725}& \multirow{2}{*}{1.830}& \multirow{2}{*}{1.908}\\
				& & & & & & & & \\
				\hline
				\multirow{2}{*}{\textbf{Only a single match}}& \multirow{2}{*}{1.266}& \multirow{2}{*}{1.325}& \multirow{2}{*}{1.400}& \multirow{2}{*}{1.488}& \multirow{2}{*}{1.582}& \multirow{2}{*}{1.671}& \multirow{2}{*}{1.750}& \multirow{2}{*}{1.805}\\
				& & & & & & & & \\
				\hline
				\textbf{Attenuation ratio}& 0.86\%& 1.12\%& 1.34\%& 1.59\%& 2.04\%& 3.13\%& 4.37\%& 5.39\%\\
				\hline
			\end{tabular}
			\begin{tablenotes}
				\item $^{\mathrm{a}}$ The compression ratio is calculated by $\frac{\textit{average size of all original files in the corpus}}{\textit{average size of all compressed files}}$. 
				\item $^{\mathrm{b}}$ In our implementation, each data block is treated as independent. We set the input buffer size to be 64KB and implement a parallelism level of 8.
			\end{tablenotes}
		\end{threeparttable}
		}
		\label{tab1}
	\end{center}
	\vspace{-2mm}
\end{table}

As shown in Table~\ref{tab1}, the compression ratio suffers from a modest attenuation ranging from 0.86\% to 5.39\% across various numbers of hash table entries, which can be considered negligible  when contrasted with the throughput enhancement achieved by aligning the actual parallelism with the theoretical parallelism.

\subsection{Elimination of the Loop Limitations}

Enhancing the frequency of the circuit is one of the most straightforward and effective ways to boost throughput. The underlying cause of the frequency bottleneck in contemporary parallel architectures\cite{bartik2019design,benevs2019high,liu2022hybridc}, as extensively illustrated in Section II-B, is the presence of a loop within the architecture. In order to break this loop, it is essential to resolve the time dependency of the address signal during the extended match stage.

In this paper, we aim to establish an upper boundary for the maximum match length, ensuring that the search for the maximum match length can be accomplished within a single clock cycle. By this approach, we no longer need to rely on the match results of the current cycle to determine the address for the next cycle, thus effectively eliminating the temporal  dependency and breaking the loop within the architecture. 

Furthermore, this operation ensures predictable output delay. By adopting the proposed strategy, the maximum match length information can be promptly obtained within a single clock cycle during the extended match stage, immediately following the discovery of the first match. This attribute distinguishes our approach from the original extended match stage, which exhibited unpredictable delay due to its reliance on the size of the match length.

\begin{figure}[tb]
	\centerline{\includegraphics[width=\linewidth]{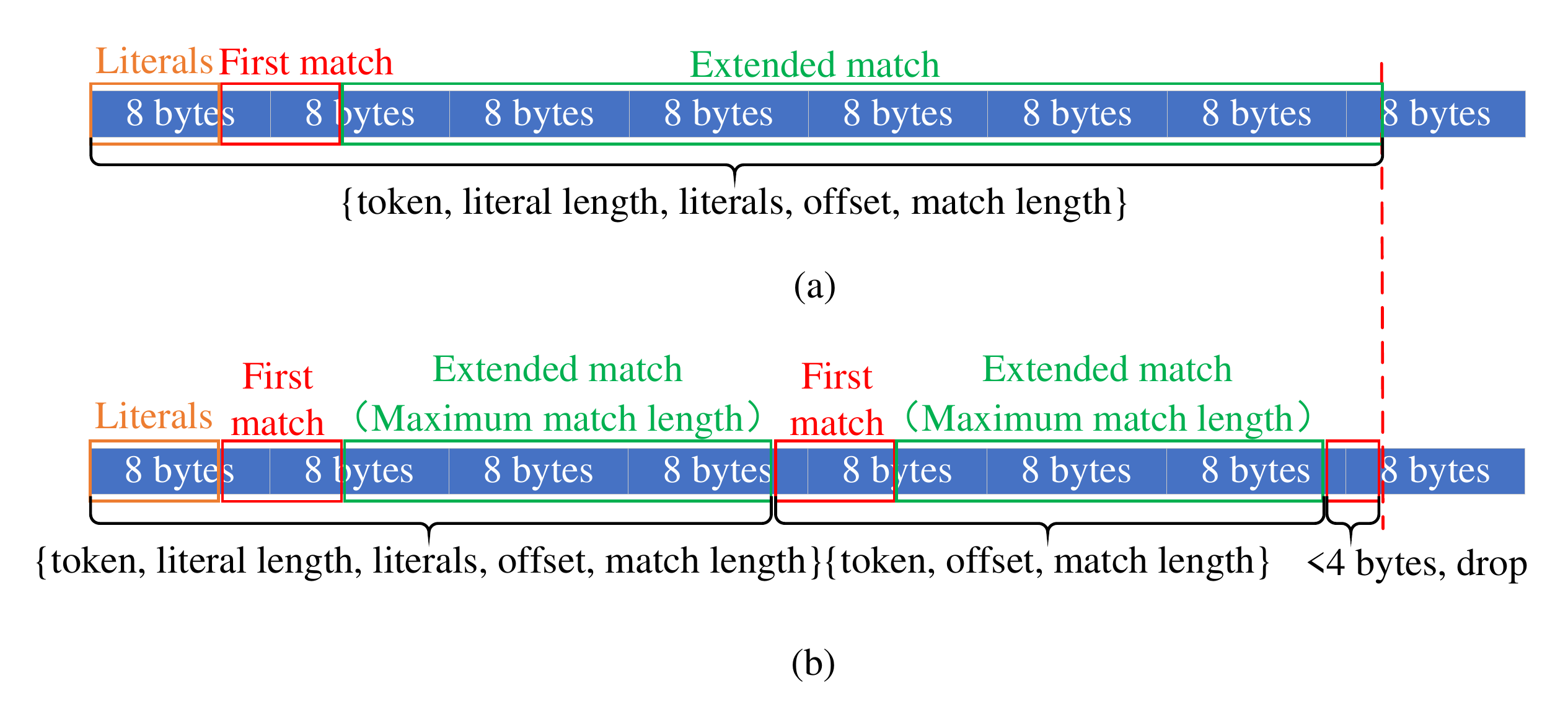}}
	\caption{(a) Original vs (b) Modified extended match stage.}
	\label{fig4}
	\vspace{-2mm}
\end{figure}

The disadvantage of establishing a maximum match length limit is apparent. As clearly depicted in Fig.~\ref{fig4}, this approach may result in the fragmentation of a long match into multiple shorter ones. Additionally, any small match at the end, measuring less than 4 bytes, will be discarded. This is due to the output format of the LZ4 algorithm, which requires a minimum match length of 4 bytes.

\begin{table}[t]
	\caption{Compression Ratio$^{\mathrm{a}}$ of Modified Extended Match Stage on the Calgary Corpus}
	\vspace{-4mm}
	\begin{center}
		\resizebox{\linewidth}{!}{
			\begin{threeparttable}
			\renewcommand{\arraystretch}{1.5} 
			\begin{tabular}{|c|c|c|c|c|c|}
				\hline
				\textbf{Number of entries}&\multicolumn{5}{|c|}{\textbf{Maximum match length}} \\
				\cline{2-6}
				\textbf{in the hash table$^{\mathrm{b}}$}& \textbf{Not Limit}& \textbf{Limit to 12}& \textbf{Limit to 20}&
				\textbf{Limit to 36}& \textbf{Limit to 68} \\
				\hline
				\textbf{64}& 1.277& 1.188& 1.204& 1.220& 1.238 \\
				\hline
				\textbf{128}& 1.340& 1.235& 1.255& 1.275& 1.295 \\
				\hline
				\textbf{256}& 1.419& 1.292& 1.317& 1.341& 1.366 \\
				\hline
				\textbf{512}& 1.512& 1.362& 1.394& 1.423& 1.451 \\
				\hline
				\textbf{1024}& 1.615& 1.439& 1.478& 1.511& 1.545 \\
				\hline
				\textbf{2048}& 1.725& 1.518& 1.564& 1.602& 1.640 \\
				\hline
				\textbf{4096}& 1.830& 1.591& 1.643& 1.685& 1.727 \\
				\hline
				\textbf{8192}& 1.908& 1.649& 1.705& 1.751& 1.797 \\
				\hline
			\end{tabular}
			\begin{tablenotes}
				\item $^{\mathrm{a}}$ The compression ratio is calculated by $\frac{\textit{average size of all original files in the corpus}}{\textit{average size of all compressed files}}$. 
				\item $^{\mathrm{b}}$  In our implementation, each data block is treated as independent. We set the input buffer size to be 64KB and implement a parallelism level of 8.
			\end{tablenotes}
			\end{threeparttable}
		}
		\label{tab2}
	\end{center}
\end{table}

To quantitatively assess the impact of this measure on the compression ratio, we conduct tests utilizing the Calgary corpus. As depicted in Table~\ref{tab2}, as the maximum match length limit increases, the decrease in compression ratio gradually wanes. 

However, setting an excessively high maximum match length limit is not always optimal. As the limit increases, the hardware resources required for match comparison gradually expand. Consequently, to achieve a balance between compression ratio and hardware resource utilization, we have chosen a maximum match length limit of 36 bytes. This choice results in a compression ratio reduction of about 4.46\% $\sim$ 8.23\%.

\subsection{Combination of the Two Schemes}

This section evaluates the change in the compression ratio when combining the two proposed schemes. The results, tested using the Calgary corpus, are presented in Table~\ref{tab3}. The compression ratio reduction ranges from 4.93\% to 11.68\% with various numbers of hash table entries. Given that throughput often takes precedence over compression ratio in numerous applications of lossless compression algorithms\cite{bartik2015lz4,bartik2020sight}, a compromise of about 10\% in compression ratio to achieve a substantial enhancement in throughput is deemed acceptable.

\begin{table}[tb]
	\caption{The Compression Ratio$^{\mathrm{a}}$ of the Combined Scheme on the Calgary Corpus}
	\vspace{-4mm}
	\begin{center}
		\resizebox{\linewidth}{!}{
			\begin{threeparttable}
			\renewcommand{\arraystretch}{1.5} 
			\begin{tabular}{|c|c|c|c|c|c|c|c|c|}
				\hline
				\multirow{2}{*}{\textbf{Scheme$^{\mathrm{b}}$}}&\multicolumn{8}{|c|}{\textbf{Number of entries in the hash table}} \\
				\cline{2-9} 
				& \textbf{64}& \textbf{128}& \textbf{256}& \textbf{512}& \textbf{1024}& \textbf{2048}& \textbf{4096}& \textbf{8192} \\
				\hline
				\textbf{GitHub\cite{GitHubLZ4}}& 1.277& 1.340& 1.419& 1.512& 1.615& 1.725& 1.830& 1.908\\
				\hline
				\textbf{Combination$^{\mathrm{c}}$}& 1.214& 1.267& 1.331& 1.409& 1.491& 1.569& 1.638& 1.685\\
				\hline
				\textbf{Attenuation ratio}& 4.93\%& 5.45\%& 6.20\%& 6.81\%& 7.68\%& 9.04\%& 10.49\%& 11.68\%\\
				\hline
			\end{tabular}
			\begin{tablenotes}
				\item $^{\mathrm{a}}$ The compression ratio is calculated by $\frac{\textit{average size of all original files in the corpus}}{\textit{average size of all compressed files}}$. 
				\item $^{\mathrm{b}}$ In our implementation, each data block is treated as independent. We set the input buffer size to be 64KB and implement a parallelism level of 8.
				\item $^{\mathrm{c}}$ Here we limit the maximum match length to 36 bytes.
			\end{tablenotes}
			\end{threeparttable}
		}
		\label{tab3}
	\end{center}
\end{table}

\section{Hardware Architecture and Implementation Results}

\begin{figure*}[htbp]
	\centerline{\includegraphics[width=\linewidth]{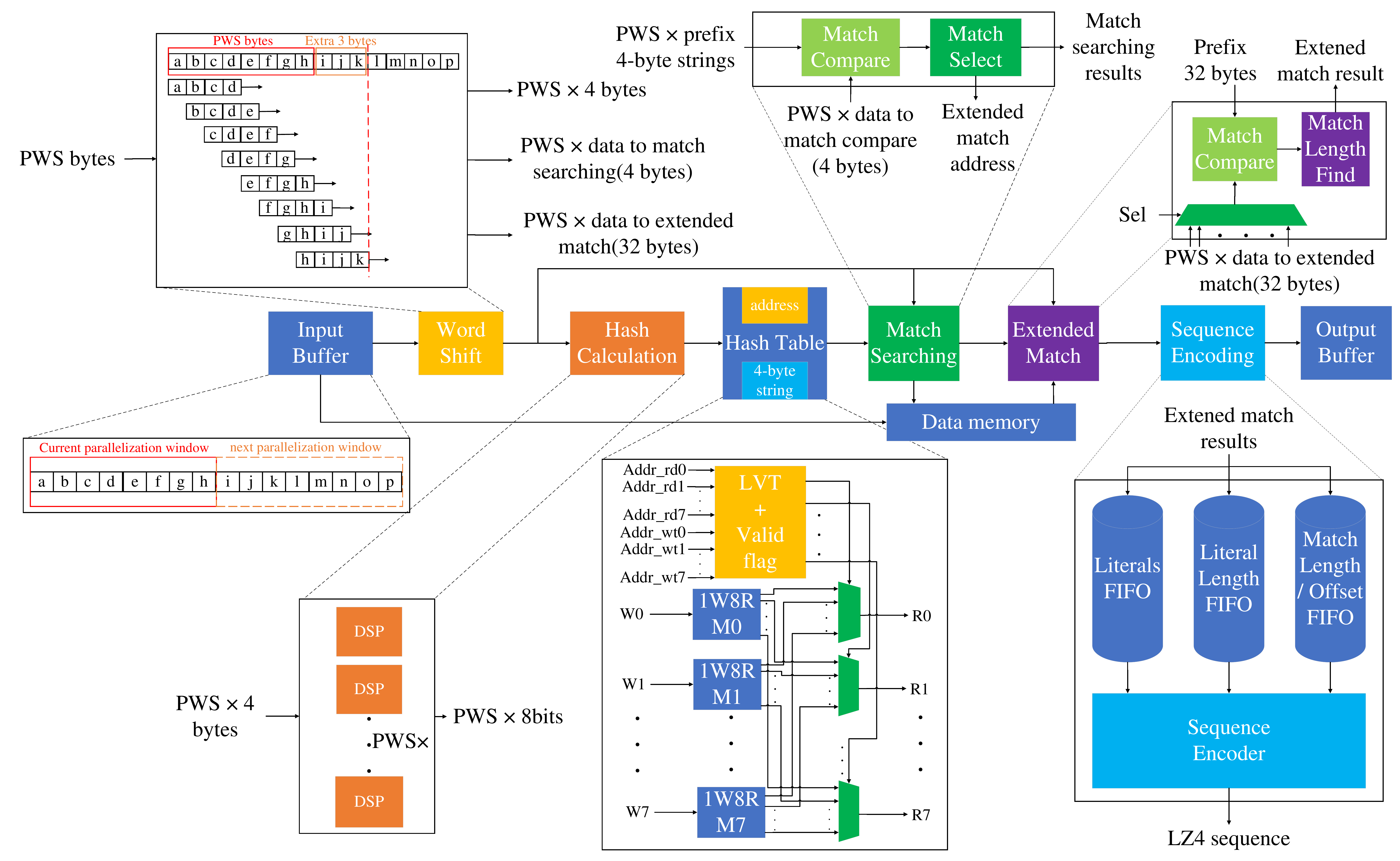}}
	\caption{The detailed architecture of the parallel compression kernel.}
	\label{fig5}
\end{figure*}

In this section, we introduce the detailed architecture of the parallel compression kernel, as shown in Fig.~\ref{fig5}. Furthermore, we evaluate the hardware resource utilization of the proposed solutions on the FPGA platform and compare it with the implementation results of prior works.

\subsection{Hardware Architecture}

The parallel compression kernel consists of several modules, all designed to facilitate multi-byte parallel processing.

\paragraph{Input Buffer} The input buffer releases a continuous stream of data, consisting of \textit{PWS} bytes per clock cycle. As depicted in the left portion of Fig.~\ref{fig5}, the processing window smoothly slides forward by \textit{PWS} bytes, generating a fresh set of \textit{PWS} bytes as output in each subsequent cycle.
Besides, the LZ4 format limits the maximum input data size to 64 KB, making it compatible with the Level-1 cache of modern processors. For convenience, here we also take the maximum value of the input buffer as 64KB.

\paragraph{Word Shift}

In this module, the input is parallelized into \textit{PWS} input strings. As depicted in the upper-left corner of Fig.~\ref{fig5}, each of these \textit{PWS} input strings is formed by assembling \textit{PWS} + 3 bytes.

\paragraph{Hash Calculation} During each cycle, the Hash Calculation module accepts \textit{PWS} strings from the Word Shift module and produces their corresponding hash values for transmission to the subsequent module. As we have mentioned in Section II-B, DSPs are employed for executing the \textit{PWS}-way Fibonacci hash computation.

\paragraph{Hash Table} This module features a parallel hash table that facilitates the simultaneous updating of \textit{PWS} records.
The Live Value Table (LVT) was chosen as the method for constructing a multi-port memory over an alternative approach that utilized an XOR technique to access the most recent value\cite{bartik2019design,benevs2019high}. 
Multi-banking is another prevalent method employed to construct multi-port memory, minimizing the utilization of BRAM resources\cite{liu2022hybridc}. Nonetheless, it comes with a notable drawback that bank conflicts may result in considerable compression ratio attenuation. 
Consequently, we opt for an architecture based on LVT to realize the multi-port hash table, as depicted in detail on the bottom of Fig.~\ref{fig5}. 

The shared hash table contains 256 items as \cite{bartik2019design,benevs2019high} did, each composed of an input buffer pointer (up to 16 bits) and a 4-byte string. Differing from the LZ4 software implementation algorithm available on GitHub\cite{GitHubLZ4}, which solely stores address pointers in the hash table, our incorporation of an additional 4-byte data element serves the purpose of addressing potential hash collisions. In the LVT scheme, the LVT also serves to store validity flags\cite{lee2017design}, eliminating the need to clear the dictionary between input data blocks.

\paragraph{Match Searching} In this module, the match result is confirmed by comparing the current and candidate strings. To determine the longest match, we select the earliest candidate among \textit{PWS} potential match candidates, prioritizing the input string located near the beginning of the parallelization window.

\paragraph{Extended Match} Based on the selection signal generated from the previous clock cycle, one of the \textit{PWS} input data is selected for comparison with the candidate string read from the data memory module. This comparison aims to identify the maximum length of match data.

\paragraph{Sequence Encoding} This module is responsible for encoding the generated compression information into an LZ4 sequence as illustrated in Fig.~\ref{fig1}.

\paragraph{Output Buffer} In previous studies\cite{bartik2015lz4}, a comprehensive analysis of the output buffer size has been conducted. For simplicity, here we opt to set the output buffer size to be the same as that of the input buffer.

\subsection{Implementation Results}

We implement our hardware architecture on an FPGA to build a high-throughput compression kernel. The selected platform is the Xilinx Kintex-7 XC7K325TFBG676-3 as \cite{liu2018data} did.

Given that in previous works\cite{bartik2019design,benevs2019high,liu2022hybridc}, the \textit{PWS} for the LZ4 parallel architecture has been consistently set to 8 bytes, we have also chosen to set our \textit{PWS} to 8 bytes for a fair comparison. The implementation results is compared with other FPGA implementations of the LZ4 algorithm. Table~\ref{tab4} presents a comparison of the designs in terms of frequency, resource utilization, and throughput. The respective implementation platforms (FPGAs) exhibit comparability in terms of resources and/or frequency due to their similarities. 

\begin{table}[t]\large
	\caption{The Implementation Results of Different LZ4 Acceleration Kernels}
	\vspace{-4mm}
	\begin{center}
		\resizebox{\linewidth}{!}{
			\begin{threeparttable}
			\renewcommand{\arraystretch}{1.5} 
			\begin{tabular}{|c|c|c|c|c|c|c|c|}
				\hline
				\multicolumn{2}{|c|}{\textbf{Scheme}}& \textbf{Bartik's\cite{bartik2015lz4}}& \textbf{MLZ4C\cite{liu2018data}}& \textbf{Benes'\cite{benevs2019high}}& \textbf{Xilinx\cite{XilinxLZ4}} & \textbf{HybriDC\cite{liu2022hybridc}}& \textbf{Ours} \\
				\hline 
				\multicolumn{2}{|c|}{\textbf{Intra-core parallelism}}& $\times$& $\times$& \checkmark& $\times$& \checkmark& \checkmark\\
				\hline
				\multicolumn{2}{|c|}{\textbf{FPGA Chip}}& XC7A100T& XC7K325T& XCKU040& XCU200& 10AX048H2& XC7K325T\\
				\hline
				\multicolumn{2}{|c|}{\textbf{FPGA Technology}}& 28nm& 28nm& 20nm& 16nm& 20nm& 28nm\\
				\hline
				& \textbf{LUT slices}& 764& 573& 14076& 3000& 12336& 10863\\
				\cline{2-8}
				\multirow{2}{*}{\textbf{Resource}}& \textbf{FF slices}& 375& 937& 2803& 3500& 4081& 3378\\
				\cline{2-8}
				& \textbf{DSPs}& 3& 4& 32& NA& 16& 32\\ 
				\cline{2-8}
				& \textbf{BRAM (Kbits)$^{\mathrm{a}}$} &612& 2484& 2952& 1908& 1460&  2880\\
				\hline
				\multicolumn{2}{|c|}{\textbf{Frequency (MHz)}}& 146& 240& 156.25& 300& 100& 251.57\\
				\hline
				\multicolumn{2}{|c|}{\textbf{Throughput (Gbps)}}& NA& 1.92& 6.08& 2.32& 4.50& 16.10\\
				\hline
				\multicolumn{2}{|c|}{\textbf{Throughput/slices$^{\mathrm{b}}$}}& \multirow{2}{*}{NA}& \multirow{2}{*}{1.272}& \multirow{2}{*}{0.360}& \multirow{2}{*}{0.357}& \multirow{2}{*}{0.274}& \multirow{2}{*}{1.131}\\
				\multicolumn{2}{|c|}{\textbf{(Gbps/K slices)}}& & & & & & \\
				\hline
			\end{tabular}
			\begin{tablenotes}
				\item $^{\mathrm{a}}$ Due to variations in the size of block RAMs across different platforms, we standardized the comparison of on-chip memory usage by converting the number of BRAMs into a consistent capacity measurement.
				\item $^{\mathrm{b}}$ Summation of LUT and FF slices. 
			\end{tablenotes}
			\end{threeparttable}
		}
		\label{tab4}
	\end{center}
	\vspace{-2mm}
\end{table}

Table~\ref{tab4} demonstrates that despite employing the worst technology conditions, our single-kernel parallel architecture
achieves a frequency of 251.51 MHz, which surpasses the highest frequency of the current parallel architecture (Benes\cite{benevs2019high}) by 1.610$\times$, and trails only behind the serial architecture (Xilinx\cite{XilinxLZ4}) due to the parallel implementation and technological reasons.
As a result of this enhanced frequency, our architecture boasts the maximum throughput among all existing LZ4 architectures, reaching a peak value of 16.10 Gbps. This throughput is an impressive 2.648$\times$ higher than the current leading throughput reported by Benes\cite{benevs2019high}.

In terms of hardware resources, our design also holds advantages. By imposing two limiting conditions, we eliminate the feedback control loops from the original circuit, resulting in a decrease in LUT and FF resources. 
As can be seen from Table~\ref{tab4}, disregarding  the MLZ4C algorithm\cite{liu2018data}, which is incompatible with the official LZ4 tool due to alterations in the LZ4 compression format, our design boasts an area efficiency ratio of 1.131 Gbps/K slices, notably surpassing other architectures, including Benes\cite{benevs2019high}, Xilinx\cite{XilinxLZ4}, and HybriDC\cite{liu2022hybridc}.

\section{Conclusion}

This paper introduces a single-kernel parallel architecture for LZ4 compression algorithm. We propose two methods to break the throughput bottleneck existing in current architectures. First, we restrict each parallelization window to only a single match to enhance the actual parallelism. Second, the maximum match length is limited to break feedback loops in the architecture and thus improve the frequency. Both methods come at the expense of sacrificing a certain degree of compression ratio.
The implementation on FPGA shows that the throughput reaches a maximum of 16.10 Gbps, surpassing the highest throughput reported for existing LZ4 architectures by 2.648$\times$.

\section{Acknowledgment}
This work was supported in part by the National Key R\&D Program of China under Grant 2022YFB4400604, and in part by Shenzhen Science and Technology Program (Grant No. RCBS20231211090610015)  (Corresponding authors: Suwen Song; Zhongfeng Wang.)

\bibliographystyle{IEEEtran}
\bibliography{Reference}

\end{document}